\documentclass[12pt,preprint]{aastex}
\newcommand{\ssst}{\scriptscriptstyle}
\newcommand{\E}[1]{\times 10^{#1}}

\newcommand{\RA}[3]{{#1}^{{\rm h}}{#2}^{{\rm m}}{#3}^{{\rm s}}}
\newcommand{\Dec}[3]{{#1}^{\circ}{#2}'{#3}''}
\def\xs{{DA~530}}

      \newcommand{\ps}{\,{\rm s}^{-1}}
\newcommand{\yr}{\,{\rm yr}}    \newcommand{\Msun}{M_{\odot}}
\newcommand{\cm}{\,{\rm cm}}    \newcommand{\km}{\,{\rm km}}

\newcommand{\erg}{\,{\rm ergs}}        
    \newcommand{\keV}{\,{\rm keV}}

\newcommand{\nel}{n_{e}}        \newcommand{\NH}{N_{\ssst\rm H}}

\newcommand{\Ts}{T_{s}}
         \newcommand{\vs}{v_{s}}
\newcommand{\nH}{n_{\ssst\rm H}}        \newcommand{\mH}{m_{\ssst\rm H}}

\newcommand{\rosat}{{\sl ROSAT}} 
\newcommand{\chandra}{{\sl Chandra}}

\newcommand{\du}{d_{2.2}}

\newcommand{\rmsn}{\langle\nH^2\rangle^{1/2}}
\newcommand{\HI}{H{\small I}} 

\begin{document}

\title{\chandra\ View of DA~530: A Sub-energetic Supernova Remnant with a Pulsar Wind Nebula?}

\author{
 Bing Jiang\altaffilmark{1,2},
 Yang Chen\altaffilmark{1}, and
 Q.\ Daniel Wang\altaffilmark{2}
}
\altaffiltext{1}{Department of Astronomy, Nanjing University, Nanjing 210093,
       P.R.China}
\altaffiltext{2}{Department of Astronomy, B619E-LGRT,
       University of Massachusetts, Amherst, MA01003}
\begin{abstract}
DA~530 (G93.3+6.9) is a high Galactic latitude supernova remnant with a 
well-defined shell-like radio morphology and an exceptionally low 
X-ray to radio luminosity ratio. Based on a \chandra\ ACIS observation,
we report the detection of an extended X-ray feature close to the 
center of the remnant with $5.3\sigma$ above the background within a circle
of $20''$ radius. The spectrum of this feature
 can be characterized by a power-law with the photon index $\Gamma=1.6\pm0.8$.
This feature, spatially coinciding with 
a nonthermal radio source, most likely represents a pulsar wind nebula.
We have further examined the spectrum of the diffuse X-ray emission 
from the remnant interior with a background-subtracted count rate of 
$\sim 0.06$ counts $\ps$ in 0.3-3.5 keV.
The spectrum of the emission can be described by
a thermal plasma with a temperature of $\sim$ 0.3--0.6 keV and a Si
over-abundance of $\gtrsim 7$ solar. These spectral characteristics,
together with the extremely low X-ray luminosity, suggest that
the remnant arises from a supernova with an anomalously low 
mechanical energy ($< 10^{50}\erg$). The centrally-filled thermal X-ray 
emission of the remnant may indicate an early thermalization of the SN 
ejecta by the circum-stellar medium. Our results suggest that the remnant 
is likely the product of a core-collapsed SN with a progenitor mass of 
8-12 M$_\odot$. Similar remnants are probably common in
the Galaxy, but have rarely been studied.
\end{abstract}


\section{Introduction}
X-ray observations of Galactic supernova remnants (SNRs) help us to explore 
how stars end their lives and what the end-products are. Existing 
studies have concentrated on X-ray-bright SNRs, which
tend to be energetic and to be located in relatively dense interstellar 
environments. The conclusions we drew from such studies may thus be biased. 
Indeed, recent observations of supernovae (SNe) show a broad range of 
explosion characteristics; some SNe appear to be substantially sub-luminous with 
very small expansion velocities of the ejecta, indicating low explosion energies
(e.g., 1997D-like SNe, Zampieri et~al. 2003, Pastorello et~al. 2004; 
the 2002cx-like SNe, Jha et~al. 2006). Naturally, one expects that
such SNe should lead to sub-energetic SNRs.
Therefore, from the study of nearby sub-energetic SNRs, we may learn about
the explosion processes and their dependence on progenitor stellar masses 
and binary properties, for example. 

DA~530~(G93.3+6.9) is probably the best known example of a sub-energetic
SNR. It has a well-defined shell-like morphology in radio 
(Roger \& Costain 1976; Landecker et~al. 1999, hereafter L99) and is extremely 
faint in X-ray (L99). It is one of the rare SNRs observed at high Galactic 
latitudes: only 7 of 265 Galactic SNRs have $|b| > 6^\circ$ in the Green 
Catalog~\footnote{Green D. A., 2006, `A Catalogue of Galactic Supernova Remnants 
(2006 April version)', Astrophysics Group, Cavendish Laboratory, Cambridge, 
United Kingdom (available at ``http://www.mrao.cam.ac.uk/surveys/snrs/'').}; 
the other 6 are all  well-known nearby SNRs, 
typically with large angular sizes. 
X-ray emission from DA 530 was first reported by L99, based on 
an 8-ksec \rosat\ PSPC observation. This observation shows that the
remnant has a centrally-filled X-ray morphology and a total
luminosity of $1.1\times10^{32}$ ergs/s in the 0.12-2.4 keV band (assuming a 
distance of $3.5$ kpc). But the very limited counting 
statistics and energy/spatial resolutions of the observation
did not allow for any tight constraints on the thermal properties of the 
SNR. Nevertheless, L99 suggested that DA~530 is probably created by a 
sub-energetic Type Ia SN in a very tenuous ambient environment. 
The total estimated 
X-ray-emitting gas mass is only about 4$\Msun$. 
Based on 21-cm \HI\ data, Foster and Routledge 
(2003) later improved the distance estimate to be $2.2\pm 0.5$ kpc 
(adopted hereafter), which would reduce the X-ray luminosity and the gas mass.
L99 further argued that the low ambient density is probably a result of a 
preexisting stellar wind bubble indicated by 
the presence of a ``necklace''-like \HI\ shell located at the outer boundary 
of the radio continuum shell. A Type Ia SN scenario is proposed because
of the remnant's relatively large vertical height (264~pc) above 
the Galactic plane; but paradoxically, the presence of the wind bubble 
is indicative of a massive progenitor (L99).
DA~530 is so far not detected in infrared and optical, consistent with 
a relatively low density environment expected at this vertical height.
Kaplan et~al.\ (2004) did not find any suitable candidates for
a potential stellar remnant of DA530 from a \chandra\ ACIS-I observation
and optical/IR follow-ups. No pulsed emission has been detected in DA~530 
(Lorimer, Lyne \& Camilo 1998).

In this paper, we report a re-analysis of the same \chandra\ observation
as obtained by Kaplan et~al. (2004). Our initial motivation is to understand
why DA530 is so faint in its X-ray emission, as shown in the 
\rosat\ study. 
If indeed a remnant of a Type~Ia~SN, it is expected to show
enhanced iron lines in the \chandra\ X-ray spectrum.
While we find no indication for this enhancement, our analysis 
(\S~2) leads to 
several quite unexpected results, which are presented in \S~3. In \S~4,
we discuss the implications of these results on the nature of the SN and its
environment. We quote statistical error bars at the 90\% confidence 
level throughout this work. 

\section{Observation and Data Reduction}\label{sec:img_01}

The \chandra\ ACIS-I observation of DA~530 was performed 
on 17 December 2001 in a faint mode (Obs.\ ID 2808). 
The target center ($\RA{20}{52}{14}.0$, $\Dec{55}{20}{30}.0$) was placed
at the aiming point of the I3 chip.
We reprocessed the data using the \chandra\ Interactive Analysis of 
Observations (CIAO) software package (version 3.2.2),
following the ACIS data analysis guide.
The lightcurve of the observation indicates no significant background flares
and the effective exposure is 16.4 ks.
The \chandra\ field of view (FoV) is shown in Fig.~1,
in comparison with DA~530's extent in radio continuum.

For imaging analysis, counts images and exposure maps are generated in the four bands: 
$0.5-1$ (S1), $1-2$ (S2), $2-4$ (H1), and $4-8$ (H2) keV. 
We subtract the non-X-ray contribution, using the ACIS stowed background 
database~\footnote{available at http://cxc.harvard.edu/contrib/maxim/acisbg/}.
All the data are corrected for CTI and gain variations.
The \chandra\ 0.5-8 keV intensity image is shown in Fig.~2. 
Since ACIS-I chips do not cover the entire field of \xs, 
we also show \rosat\ PSPC intensity contours for comparison.

Our main interest here is the diffuse emission from the remnant's interior.
Thus we need to subtract the irrelevant point sources from the image.
The source detection by Kaplan et~al. (2004) is designed for finding
the putative neutral star with a select criteria of $\le 10$ counts and within
only $8'$ radius from the center of the remnant.
Following the procedures detailed in Wang (2004), we search for discrete 
sources over the entire FoV of the I-array
in the soft (S=S1+S2), hard (H=H1+H2) and broad 
(S+H) bands. 
The detection is based on a combined set of algorithms: 
wavelet, sliding box, and maximum likelihood centroid fitting, 
which are optimized for detecting point-like sources.
The map detection and the maximum likelihood analysis use a detection
aperture of the 70\% energy-encircled radius.
%Multiple detections with overlapping $2\sigma$
%centroid error circles are considered to be identical and the centroid
%position with the smallest error is adopted.
The threshold of the local false detection probability is $P\le 10^{-6}$,
based on the Poisson statistics.
We detect all the 18 sources listed in Kaplan et~al. (2004), plus 30 
additional ones, which are all marked in Fig.~3. 

To study the diffuse X-ray emission, we remove from the data a circle 
of 1.5 times the detection radius for each source.  
We then produce a broad band (H) intensity image 
of the diffuse X-ray emission (Fig.~3).

\section{Data Analysis and Results}\label{sec:img_02}

The PSPC image contours show a clear enhancement of the ``diffuse'' X-ray 
intensity in the field of \xs\ and exhibits no indication for any
rim-brightening associated with the outer radio shell (Fig.~2; L99). 
Using the NED/SIMBAD database, we find that
all the bright point sources around the rim are background sources (L99).
The ACIS-I image further reveals a faint large-scale diffuse 
emission in the interior (Fig.~3). 
The morphology appears to be centrally-filled with clumps of 
enhancements.
 
\subsection{Diffuse emission}
We perform a spectral analysis of the large-scale diffuse X-ray emission
associated with DA~530 using the \chandra\ observation.
We extract the on-SNR spectrum from the four sectors (radius $\sim 8.5'$) 
on the ACIS-I chips, which are projected entirely within the boundary of the 
SNR, and the off-SNR spectrum from the two rectangular regions on the ACIS-S2
and S3 chips (see Fig.~1). We subtract from the on- and off-SNR spectra 
the respective non-X-ray background contributions estimated from
the same regions.
After being adaptively grouped to achieve a $S/N > 3$ in each bin,
the two spectra %, on and off the SNR,  
are jointly fitted (see Fig.~4), in which the on-SNR sky background is 
determined by the fit to the off-SNR spectrum, which is well described by a 
power-law with a photon index $\sim 2$.
This double background subtraction method accounts for the position dependence
of the background, effective area, and energy response of the instruments.
Although the counting statistics of the data are limited,
the on-SNR spectrum with a net background-subtracted count rate of $\sim 0.06$ 
counts $\ps$ shows
a distinct Si~He$\,\alpha$ ($\sim 1.85$ keV) line.
The feature at $\sim 1.34$ keV may represent the He-like Mg~K$\alpha$ line,
but this line is too weak to be certain.

The net on-SNR spectrum of the diffuse X-ray emission is well 
characterized with a non-equilibrium ionization thermal plasma model 
({\em vpshock} in XSpec). We present in Table~1 the fit results 
with or without fixing the column density $\NH$ to $5.7\E{21}\cm^{-2}$,
an estimate of the total column density based on observations of atomic
and molecular gas in the field (L99). This independent estimate represents
an upper limit to the true X-ray absorption column density along the
sight line to DA 530, depending on its distance. If the remnant is indeed 
at the distance of $\sim 2.2$ kpc (or $\sim$300 pc above the Galactic plane), 
beyond the bulk of the neutral gas disk, the estimate should be close to the
true absorption column density. The column density from the direct fit
to the X-ray spectrum is, however, considerably smaller than the estimate,
which may indicate that the remnant is much closer than the adopted 
distance. But there are uncertainties in the X-ray data calibration
(e.g., background subtraction) and modeling. We hence present the results
based on these two different absorption column density estimations.
There is no significant evidence for an enrichment in Fe 
(a fit indicates an iron abundance close to the solar value),  
while the abundance of Si is clearly super-solar.
The large ionization parameter ($n_e t_i \sim 10^{12} \cm^{-3}$\,s)
indicates that the plasma is close to the collision ionization equilibrium.
An equilibrium ionization plasma model ({\em vmekal}) is then applied
and the fitted parameters, as well as the confidence ranges, are similar 
to those from {\em vpshock}.

\subsection{The central extended hard X-ray feature}
Fig.~3 shows a remarkable small-scale X-ray feature at
${\rm RA}=\RA{20}{52}{14}$, ${\rm Dec.}=\Dec{55}{17}{22}$,
near the center of the remnant. This central feature is present chiefly in 
the range of $1-4$ keV. The radial intensity profile of the feature 
(Fig.~5) clearly shows that it is an 
extended source. A fit with an exponential function
($\propto e^{-r/r_0}$) plus a local background gives a scale-length of 
$6^{\prime\prime}\pm3^{\prime\prime}$.
We further extract a spectrum of the feature (Fig.~6) in 
a circle of $20''$ radius (marked with ``S'' in Fig.~1),
which includes about 96\% net counts according to the best-fit exponential 
function. The total number of counts in 0.3-7 keV band is 50, compared to
a local background contribution of 24 counts, estimated from 
an annulus between $40''$ to $80''$ around the feature.
Therefore, the significance of the feature above the local background is 
5.3$\sigma$. Accounting for the overall FoV 
($17^\prime\times17^\prime$) of the ACIS-I and the size 
($20^{\prime\prime}$ radius) of the feature, the statistical probability 
for it to be due to the random fluctuation of the background is on 
the order of $10^{-5}$. Of course, there are systematic effects (e.g., the
non-uniformity of the diffuse emission; see \S 4.1 for further discussion), 
which cannot be adequately quantified here. Indeed, there are a few other 
peaks in Fig.~3. But they are all at large off axis angles and may at 
least partly due to confusions among adjacent weak sources because of the 
large PSF. No peak in the central region has a significance close to the above 
feature or is associated with a radio source (\S~4.1). We conclude that 
the feature is most likely real and is truly extended.

With the very limited counting statistics, the spectrum of the feature 
(after the local background subtraction) 
can be well fitted with
a variety of models, assuming the hydrogen column density of 
$N_H = 5.7\E{21}\cm^{-2}$. For example,
the power-law model gives a photon index of $1.6\pm 0.8$
and an unabsorbed $0.5-10\keV$ flux of $\sim 3.6\E{-14}\,\erg\cm^{-2}\ps$.
Alternatively, an optically-thin thermal plasma would indicate a 
temperature $\gtrsim 2$ keV, substantially higher than that found for 
the surrounding diffuse X-ray emission ($\sim 0.5$ keV; Table~1).
With a lower $N_H$ value assumed (e.g. $1.5\E{21}\cm^{-2}$), the temperature
would be even higher, while the photon index in the power-law model 
would be smaller  ($1.1\pm0.7$) with nearly the same unabsorbed $0.5-10\keV$ flux.

The central extended hard X-ray feature is not seen in the \rosat\ PSPC
observation, which had a low spatial resolution and was sensitive only to
X-rays below $\sim2.4\keV$. The feature was also not detected in the ACIS
observation, based on the CIAO {\em wavdetect} program with a maximum 
wavelet scale of $8''$ (Kaplan et~al. 2004). However, with
larger scales of 32, 64, 128 pixels, we do detect the feature, using the
same program. 

\section{Discussion}

We now discuss the implications of the above results, together with those
from previous studies.

\subsection{Pulsar Wind Nebula Candidate}\label{sec:pwn}

We have shown the evidence for the presence of
a central extended X-ray feature in DA 530. This feature is characterized 
by an angular extent of $\sim 40''$ and a hard spectrum. 
This hard X-ray feature, with its small extent, could conceivably represent a 
distant ($z \gtrsim 1$) cluster of galaxies projected in the field. 
However, such clusters are very rare (only $\sim 10$ X-ray-emitting
clusters have been detected at $z>1$; Stanford et~al. 2006). 
The observed spectrum (Fig.~6) also seems to be a  
too hard to be consistent with the expected relatively steep (after the redshift)
thermal emission from the intracluster medium.
Alternatively, the feature might just be an 
enhancement of the X-ray emission from \xs\, due to a density increase, 
for example. But assuming a rough pressure balance, one would expect that 
such density enhancement tends to have a lower temperature than the 
surrounding, inconsistent with the hard spectrum of the feature (\S 3.2).

We propose that the hard X-ray feature most likely represents a pulsar wind nebula (PWN).
This interpretation naturally explains the location near the center of \xs, 
the spatial extension, and the apparently hard spectrum. 
In this case, the X-ray emission represents the synchrotron radiation from 
relativistic electrons/positrons in the nebula. The observed spectrum
in particular, characterized by a power law with the photon 
index $\sim1.6\pm 0.8$, is consistent with the typical range 
$1.3-2.3$ for a PWN (Gotthelf 2003; Gaensler \& Slane 2006). 

The PWN scenario of the X-ray feature is further supported by its positional
coincidence with an extended radio source. It is detected in the 
1.4 GHz NRAO VLA Sky Survey (Condon et~al. 1998; Fig.~7) and is listed as 
``J205213+551721''
in the online database (http://www.cv.nrao.edu/nvss/NVSSlist.shtml). 
Fig.~7 also includes a radio image of \xs, 
extracted from the 325 MHz Westerbork Northern Sky Survey 
(Rengelink et~al. 1997) with a radio intensity enhancement apparent 
at the source position. The source is also present in the 
408 MHz and 1.42 GHz continuum maps of \xs\ (L99).
The radio fluxes of the source are $40\pm6$ mJy at 325 MHz and $15.0\pm1.4$ mJy 
at 1.4 GHz, which give an approximate estimate of the power law spectral slope 
as $\sim 0.7$, comparable to the typical value $\sim 0.5$ of known PWNe. But 
sensitive radio observations with a higher spatial resolution is needed to
further the test of the PWN scenario. No obvious optical or near-IR 
counterpart is found within a few arc-seconds around the centroid of 
the X-ray feature. This is expected for the PWN interpretation. 
The extension of the radio source 
($\sim 36.8^{\prime\prime}\times30.6^{\prime\prime}$, deconvolved FWHM 
source sizes in NVSS Source catalog) 
is larger than that of
the X-ray feature. This is consistent with the short synchrotron cooling
time scale of the X-ray-emitting particles.

Of course, the PWN nature of the X-ray feature and the radio peak can be 
established only when pulsed radiation is detected. While such radiation 
has not been observed, either due to the limited sensitivity
of the existed radio searches (Lorimer et~al. 1998) or to the possibility
that the pulsed radiation is beamed away from the Earth, we proceed here 
with the hypothesis that the X-ray/radio feature represents a PWN in \xs. 

Using the empirical relationship given for PWNe (Seward \& Wang 1988)
and the 0.2-4 keV luminosity of the feature
($L_x\sim 1.6\E{31}\,d^2_{2.2 {\rm~kpc}}$\erg\ s$^{-1}$ obtained from the 
spectral analysis),
we infer the spin-down power of the putative pulsar as 
$\dot{E}\sim 2.4\E{34}\,d_{2.2 {\rm~kpc}}^{1.4}\,\erg\,s^{-1}$.
According to the estimated $F_{PWN}/F_{PSR}$ of 17 pulsars in 2-10 keV  
(Gotthelf 2004) the ratio seems to be around 1 among the pulsars with 
low $\dot{E}$. Then the X-ray luminosity for the pulsar will be 
$\sim10^{31}\erg\ps$.
The deduced values, both the X-ray luminosity and $\dot{E}$, are much smaller 
than that of other PWN/PSR systems (Gotthelf 2003, 2004).
However, the pulsar PSR B1929+10 has an even smaller luminosity ($L_x\sim
10^{30}{\rm~ergs~s^{-1}}$) and $\dot{E} (\sim 4\E{33} {\rm~ergs~s^{-1}}$) with 
its faint nebula ($L_x\sim 1.3\E{30} {\rm~ergs~s^{-1}}$) (Wang, Li \& Begelman 
1993; Yancopoulos, Hamilton \& Helfand 1994; Becker et~al. 2006).

\subsection{Energetics of the SNR}\label{sec:Ia}

From the results on the diffuse X-ray emission associated with \xs\ (Table~1), 
we estimate various physical parameters of the hot gas enclosed in \xs\ 
(Table~2). The estimation of the total mass ($M$) and mean density ($\rmsn$) 
of the X-ray-emitting plasma is based on the volume emission 
measure (Table~1) and the remnant's radio size ($r_s \sim 14'\sim9\,\du$~pc). 
A volume filling factor ($f$) of the hot gas is included.
The mass estimation adopts a 
correction factor of 2, accounting for the partial spatial field
of the on-SNR spectrum within the radio shell. This correction 
assumes no particular enhancement associated with the
rims of the remnant, as indicated by the PSPC observation. 
We also estimate the total Si mass (M$_{Si}$) and thermal energy ($E_{th}$)
of the hot gas, using the abundance and 
temperature values from the spectral fits. 

We may further infer the age and explosion energy of \xs\ by applying 
the canonical Sedov (1959) blastwave model.  
If the temperature $T_x$ measured from the {\em vpshock} model 
is adopted as the post-shock temperature $T_s$, we can then 
estimate the blastwave velocity $\vs=(16k\Ts/3\bar{\mu}\mH)^{1/2}$, 
the dynamic age $t=2r_s/5v_s$ and the explosion energy 
$E=(25/4\xi)\rho_1 v_s^2 r_s^3 $ (see Table~2), 
where the mean atomic weight $\bar{\mu}=0.61$, $\mH$ is the mass 
of a hydrogen atom, $\xi=2.026$ and $\rho_1$ is the pre-shock medium mass 
density which is assumed to approximately be the mean swept-up gas density.
The low temperature ($0.3-0.6$ keV) of the gas is 
consistent with the inferred relatively large age ($\sim 5000-7000\yr$) 
of the remnant. 
The explosion energy is very low $\sim 1-4\E{49}\,d_{2.2}^{2.5}f^{-1/2}\erg$,  
about two orders of magnitude lower than the nominal value ($10^{51}\erg$) 
of a SN.
Of course, $T_x \not= T_s$ in the Sedov solution; the ACIS-I field 
covers a combination of the remnant's physical interior 
and projected outer shock regions. In the Sedov phase, the inner gas is hotter 
than the outer gas (i.e., $T_x > T_s$). 
Therefore, the above estimates represent the upper limits to the thermal and 
explosion energies and the lower limit to the age.
Considering the uncertainty on distance ($2.2\pm 0.5$ kpc), 
if the remnant is closer than 2.2 kpc,
the mass, age, and energy would be even smaller than
the values we obtained here; if we take the high end of the distance
(2.7 kpc), these parameters would be scaled up a bit,at most with a
factor of $\sim1.7$, which would not dramatically affect the final results.
Furthermore, the forward shock may have already entered a radiative phase
(i.e., $T_s \lesssim T_c=6\times 10^5$ K), consistent with the lack of 
significant rim-brightened X-ray emission (L99; \S~3). This requires
the pre-shock density $n_1 > \left(3/100\right) \left(
\bar{\mu}/1.4 kT_c\right) \left(\xi E/r_s^3\right)$, 
 i.e., $n_1 > 1.5 E_{50} \cm^{-3}$, where $E_{50}$ is the explosion energy 
in units of $10^{50}\erg$.
An age of  $\ga 1.7\E{4}$ yr is then expected in this case.
Since the preshock gas density may be similar to the mean density of the
hot gas inside the SNR ($< 0.1 \cm^{-3}$, Table 2), 
%Since the mean interstellar medium density on the Galactic plane is about 
%$1 \cm^{-3}$, 
%the density around DA530 at the relatively high Galactic latitude is 
%expected to be considerably lower than that.
$E$ should then be smaller than $10^{50}$ ergs. 
In conclusion, the SN
responsible for \xs\ appears to be sub-energetic.

\subsection{X-ray Morphology}

The X-ray results from both \rosat\ and \chandra\ suggest that
\xs\ is a so-called thermal composite/mixed-morphology SNR. 
Whereas the radio continuum emission shows a well-defined shell-like structure, 
there is little evidence for any associated enhancement in X-ray. 
This lack of the X-ray emission
associated with the forward shock is consistent with
the large ionization timescale of the remnant interior and low temperature,
as discussed above. 

The formation mechanisms for thermal composite SNRs in general are still very
much uncertain (e.g., Shelton et~al. 2004; Chen et~al.\ 2004; 
Kawasaki et~al. 2006). 
These SNRs most likely represent a heterogeneous 
population, depending on the ambient medium density distribution as well
as the remnant ages. Known thermal composites tend to be in dense environments 
(see Table 7 in Lazendic \& Slane 2006). 
This may partly be due to observational selection effects, because 
such remnants tend to be brighter. With more sensitive observations, 
fainter ones have been detected in less dense environments (Lazendic \& Slane 2006). 
None of the proposed mechanisms (e.g, evaporation of clouds, 
thermal conduction effect, and ejecta enrichment) seem to give satisfactory 
explanations for the observed properties of such thermal composites. 
This is even more difficult for \xs, which appears to represent an extreme case. 
\xs\ seems to belong to the group of thermal composites that show enhanced 
metal abundances in their interior.
The number of such remnants is now up to $\sim 50\%$ of the total thermal 
composites (Lazendic \& Slane 2006).
Detailed studies of the abundances and their distribution are needed 
to investigate their effect on the central brightness.

We find that the morphology and other X-ray properties of \xs\ can be
explained if the SN ejecta were thermalized early from the interaction with a
dense circum-stellar medium (CSM). Without this early thermalization,
the current low density would not be consistent with the large $n_e t_i$ of 
the interior hot gas (Table~1).
The required average density needs to be as high as 
$10\,\cm^{-3}$ for an age of 5000 yr. 
Such a high density can probably
only be realized in a dense CSM surrounding the progenitor. 
The required total mass of the CSM does not need to be very high (e.g., a few
$\Msun$), since the ejecta mass should be small 
as indicated by the low hot gas mass (Table~2).
Similar phenomena are also found in the LMC SNRs DEM~L238 and DEM~L249 
(Borkowski et~al. 2006). 

\subsection{Nature of the Explosion}

Let us now examine what SN progenitor may have the inferred 
properties: 1) the production of a pulsar, 2) the low mechanical energy and 
ejecta mass, and 3) the presence of a dense CSM. 
A pulsar may be produced in a core-collapsed (CC) SN of a
star with a main-sequence mass $\gtrsim 8 \Msun$, an accretion-induced
collapse (AIC) of a white dwarf, or a merger-induced
collapse (MIC) of two white dwarfs with a combined
mass greater than the Chandrasekhar mass. The latter two scenarios, though
plausible, may work only for white dwarfs with O/Ne/Mg cores and 
with fine-tuned accretion rates. The expected explosion rate is low in the 
Galaxy (e.g., Saio \& Nomoto 1985; King et~al.2001). Also no dense CSM is 
expected from the progenitor of an AIC or MIC SN. While a wind may 
be present during the accretion, it is expected to be as fast as 1000 - 1500 
${\rm~km~s^{-1}}$ (Hachisu \& Kato 2003), producing no dense CSM. 
We thus consider the massive star CC SN to be the most probable scenario for
\xs. In this case, the CSM can naturally arise from a massive slow wind 
in a post-main sequence evolutionary stage (e.g., red supergiant). Such a
CC SN typically yields a mechanical energy of $\sim 10^{51}\erg$ , 
empirically. As mentioned in \S~1, this empirical expectation may be
strongly biased. 
Type II-P SNe such as 1997D have been observed to be 
sub-energetic events (Zampieri et~al. 2003, Pastorello et~al. 2004).
There are two plausible progenitor models for them, 
a massive star (e.g. $25-40 \Msun$ for SN~1997D), which is likely 
accompanied by the formation of a black hole (Zampieri et~al. 1998),
and an intermediate-mass star ($8-12\Msun$) (Chugai \& Utrobin 2000; 
Smartt et al.\ 2004).
When examining the two models, we found that
SNRs Crab and 3C58 may have provided proper evidence 
for the nature of the DA530 case. Both the plerionic
SNRs were observed to have pre-SN mass loss or the CSM and their kinetic
energy is $\sim10^{49.5} \erg$ (Rudie \& Fesen 2007), much lower than 
the canonical $10^{51} \erg$ and similar to the value obtained in \xs.
Such subenergetic SNRs are suggested to result from the ``electron
capture'' SNe, which are thought to be the end stage of intermediate mass
stars (8--12$\Msun$), because the SNe collapse with timescales determined
by electron capture rate rather than dynamical timescale and thus the
released kinetic energy is lower (Nomoto 1987; Rudie \& Fesen 2007). 
On the other hand, recent hydrodynamic simulations incoporating
electron capture also show that CC~SNe of stars in the mass range of
$8-12\Msun$ with O-Ne-Mg cores can be substantially sub-energetic
(Kitaura et~al. 2006). Therefore, in view of the small ejecta mass
(Table~2), the pre-SN CSM, the low explosion energy, and the presence
of the PWN (hence the presence of a neutron star), we favor the low mass
CC explosion for DA~530.

The interpretation of an 8--12$\Msun$ main-sequence progenitor for the 
\xs\ SN is also
consistent with the observed pre-existing stellar wind bubble (L99). Before  
evolving to the red supergiant stage, the star is expected to undergo 
a fast stellar wind period with 
a mass loss rate of $\sim10^{-10}-10^{-7}\Msun\rm /yr$ (Snow 1982), which 
can naturally lead to the formation of a surrounding stellar wind bubble.
A wind bubble can reach a stall radius: 
$r_{\rm stall}=43.3 \left(L_{w,36}/n_0\right)^{1/2} C_{0,6}^{-3/2}$ pc, 
where $L_{w,36}=L_w/10^{36}\erg\cm^{-3}$ is the wind luminosity
in units of $10^{36}\erg\ps$, $n_0$ is the undisturbed ambient interstellar 
medium density,
and $C_{0,6}$ is the isothermal sound speed of the medium
in units of $10^6\cm\ps$ (eg., Brighenti \& D'Ercole 1994).
If $r_{\rm stall}$ is taken to be the H~\small I \normalsize ``necklace'' 
radius $\sim 9$ pc, then the stall time 
$t_{\rm stall}=2.2\E{6}\left(L_{w,36}/n_0\right)^{1/2}C_{0,6}^{-5/2}\ {\rm yr}\sim 5\E{5}$ yr (assuming $C_{0,6}=1$),
reasonably smaller than the lifetime of the progenitor.
From $L_w=\frac{1}{2}\dot{\rm M}v_w^2$, we require 
$\dot{\rm M}\sim 1.2\E{-9} \left(r_{\rm stall}/9\ \rm pc\right)^2\left(n_0/0.1 \cm^{-3}\right) \left(v_w/1000 \km\ps\right)^{-2}C_{0,6}^3$ $\Msun/\rm yr$,
where $v_w$ is the velocity of the stellar wind. 
The wind power of a 8--12 $\Msun$ main-sequence progenitor is high enough to 
generate such a bubble. 

In short, we have constructed a plausible scenario for \xs.
The progenitor could be a star with its
main-sequence progenitor mass in the range of $\sim$ 8--12 $\Msun$.
The fast wind in the main-sequence stage produces a wind bubble ($\sim9$ pc),
which is responsible for the observed HI ``necklace''-like structure.
The massive slow wind in the later red supergiant stage generates a 
dense CSM, which explains the large ionization time scale of the 
interior hot gas.
The SN apparently has a substantially low mechanical energy 
and a small ejecta mass and leaves a neutron star as the stellar remnant.

If \xs\ indeed represents a typical SNR produced by stars 
in the main-sequence mass range of $8-12 \Msun$,
we should then expect a substantial number of similar remnants in the Galaxy; 
stars in this mass range account for about one quarter of the total 
number of stars with
masses $\gtrsim 8 \Msun$ (assuming the Salpeter initial mass function) for
which CC SNe are expected. Because of the sub-energetics of these remnants, 
they are generally difficult to detect. Possible known candidates include
the Crab Nebula (Nomoto et~al. 1982) and
VRO 42.05.01 (G166.0+4.3, Burrows \& Guo 1994) at relatively high 
Galactic latitudes, 3C~58 (Rudie \& Fesen 2007), 
and those low-surface brightness SNRs at low latitudes
(Bamba et~al. 2003). Those latter SNRs show primarily hard
{\sl non-thermal} X-ray emission, which can be understood if the soft
{\sl thermal} radiation from low temperature shock-heated gas has largely
been absorbed by the intervening cool interstellar medium. 
Therefore, more detailed studies with improved data on high-latitude 
Galactic SNRs as well as those in the LMC are essential to the understanding 
of their true nature and their implications for the evolution of the Galaxy.
Meanwhile, deeper observations in both radio and X-ray are needed
for a detailed investigation of the PWN candidate of \xs.

\section{Summary and Conclusion}

We have conducted the spatial and spectral analysis 
of the X-ray faint SNR DA~530, based chiefly on a \chandra\ 
ACIS-I observation. We have detected a small-scale, extended, hard X-ray 
feature near the remnant center. Its X-ray characteristics, 
as well as the spatial coincidence with a radio continuum emission peak, 
strongly suggests that the feature is a PWN associated with \xs.

\xs\ appears to be a SNR with a well-defined radio continuum 
shell and an apparently centrally-filled thermal X-ray morphology, i.e. the 
so-called thermal composite SNR. 
The diffuse X-ray emission from the interior of the remnant is weak 
and shows spectral characteristics of an optically-thin thermal
plasma over-abundant in silicon. The inferred low mass and energy in
this plasma suggest that the explosion energy is $< 10^{50}\erg$. 
The near collisional
ionization equilibrium of the plasma further suggests that the SN ejecta
have been thermalized early in the evolution of the remnant, probably due to 
the presence of a dense CSM produced by the SN progenitor. 

The presence of the PWN and the dense CSM, as well as the low explosion energy, 
suggest a stellar progenitor with a mass probably in the range of $8-12\Msun$, 
consistent with the observed HI bubble (L99).
As such, \xs\ provides evidence for a nearby 
sub-energetic SN, which generates a pulsar and a remnant both of X-ray faint.

\acknowledgements
We thank Zhen-ru Wang, Xiangdong Li, Zhiyuan Li, Yang Su, and Jiangtao Li for  
helpful discussions and Michael D. Stage for many useful comments.
Y.C.\ acknowledges the support from NSFC grants 10673003 and 10221001,
while Q.D.W.\ acknowledges the support from NASA/CXC under the grant NAG5-6057X.

\clearpage

\begin{deluxetable}{l|cccccccccc}
\tabletypesize{\footnotesize}
\tablecaption{Spectral fit results of \xs}
\tablewidth{0pt}
\tablehead{
\colhead{parameter}  \vline & 
\colhead{free $\NH$} & \colhead{fixed $\NH$}
}
\startdata
$\NH$ ($10^{21}\cm^{-2}$) & $1.5^{+1.1}_{-0.1}$ & $5.7^{\rm a}$ \\
\\
$kT_x$ (keV) & $0.54_{-0.09}^{+0.07}$ & $0.34^{+0.17}_{-0.03}$  \\
{[Si/H]} & $13^{+9}_{-5}$ & $13^{+12}_{-6}$  \\
$n_e t_i$ ($\cm^{-3}\,{\rm s}$) & $2\E{12}(>4\E{11})$ & $6(>1)\E{11}$  \\
$\int\nel\nH\,dV/\du^{2}$ ($10^{55}\cm^{-3}$)$^{\rm b}$
 & $1.4^{+0.8}_{-0.7}$ & $9.8^{+7.0}_{-5.2}$  \\
\\
$F^{(0)}(0.5-10\keV)$ ($\erg\cm^{-2}\ps$) & $7.7\E{-13}$ & $5.1\E{-12}$ \\
$\chi^{2}/$d.o.f & 73/72 & 83/73 \\
\enddata
\tablecomments{Based on the spectral fits to the interior diffuse X-ray emission of 
\xs\ with an absorbed {\em vpshock} model. Parameter errors are quoted at the
90\% confidence.}
  \tablenotetext{a}{\phantom{0}The total hydrogen column density along the line of sight
 adopted from the observations of atomic and molecular gas (L99).}
\tablenotetext{b}{\phantom{0}The volume emission measure.}
\end{deluxetable}

\clearpage

\begin{deluxetable}{l|cccccccccc}
\tabletypesize{\footnotesize}
\tablecaption{Physical parameters of \xs}
\tablewidth{0pt}
\tablehead{
\colhead{parameter}  \vline &
\colhead{free $\NH$ (1.5)} & \colhead{fixed $\NH (5.7)$}
}

\startdata
$\rmsn\du^{1/2}f^{1/2}$ ($10^{-2}\cm^{-3}$)$^{\rm a}$ & $1.6^{+0.5}_{-0.4}$ & $4.3^{+1.5}_{-1.1}$ \\
M$\ \du^{-5/2}f^{-1/2}\ (\Msun)$ & $1.7^{+0.5}_{-0.4}$ & $4.5^{+1.6}_{-1.2}$ \\
M$_{\rm Si}\ \du^{-5/2}f^{-1/2}\ (\Msun)$ & $0.01^{+0.01}_{-0.01}$ & $0.04^{+0.04}_{-0.02}$\\
$v_s$ ($\km\ps$) & $670^{+45}_{-60}$ & $530^{+130}_{-26}$ \\
$E_{th}\du^{-2}f^{-1/2}$ (ergs) & $4.3^{+1.4}_{-1.3}\E{48}$ & $1.1^{+0.6}_{-0.2}\E{49}$\\
$E\du^{-5/2}f^{1/2}$ (ergs)$^{\rm b}$ 
& $1.1^{+0.4}_{-0.3}\E{49}$ & $2.9^{+1.5}_{-0.4}\E{49}$ \\
$t\du^{-1}$ ($\yr$)$^{\rm b}$ 
& $5200^{+460}_{-350}$ & $6600^{+300}_{-1600}$ \\
\enddata
\tablecomments{The parameters are estimated from the spectral fit results 
in Table~1. Errors are propogated from the confidence ranges of the temperature,
the abundance, and the emission measure. } 
  \tablenotetext{a}{\phantom{0}$f$ denotes the volume filling factor of the hot gas.}
  \tablenotetext{b}{\phantom{0}inferred from the Sedov modeling.}
\end{deluxetable}

\clearpage

\begin{figure}
\centerline{ {\hfil\hfil
\includegraphics[height=5.5in,angle=0, clip=]{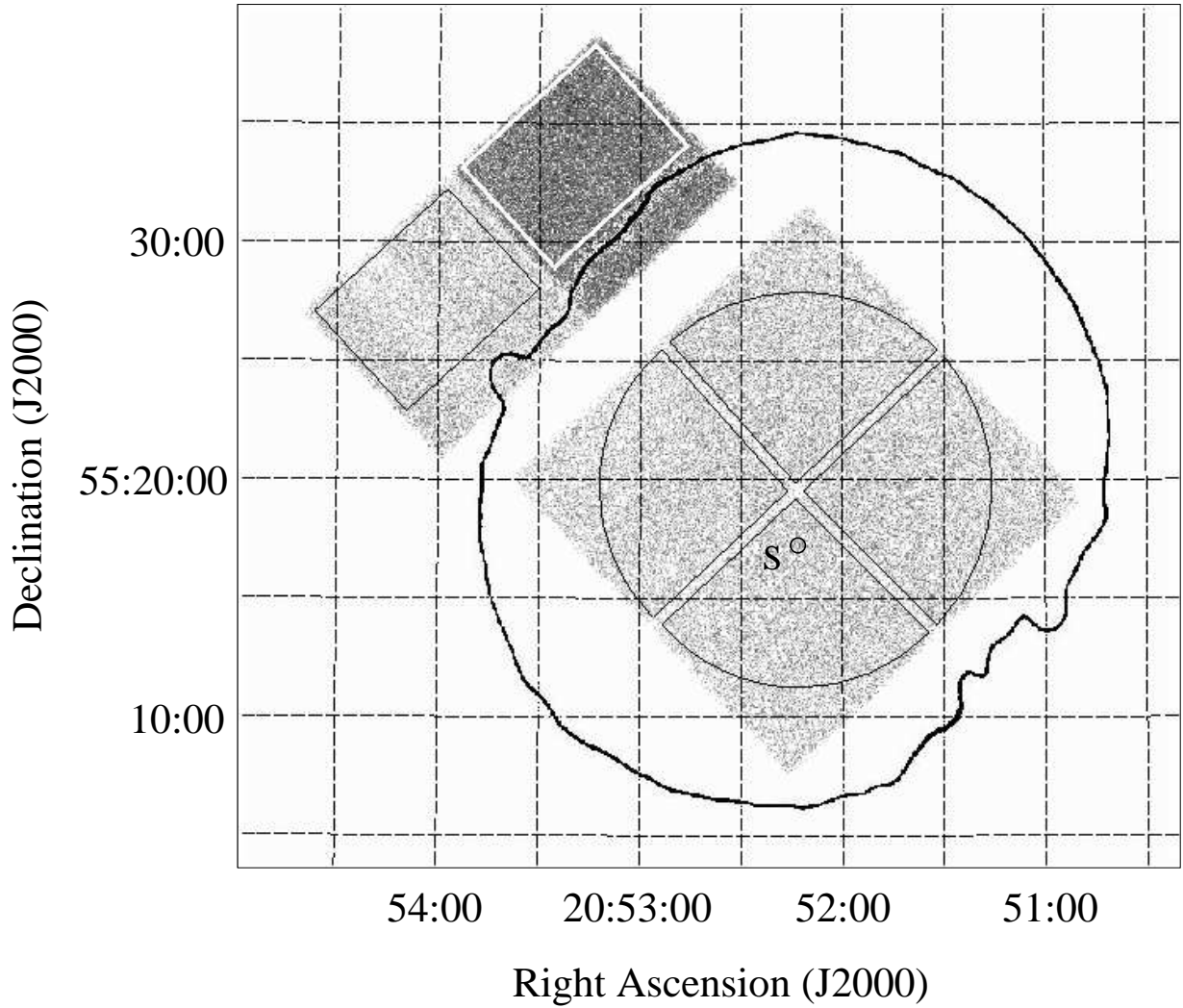}
\hfil\hfil}}
\caption{
The field of view of the \chandra\ ACIS observation of SNR~DA~530, compared with
its outer boundary as outlined by a single 1420 MHz continuum intensity 
contour at 2 mJy beam$^{-1}$  (L99).
Detected point-like sources (Fig.~3) have been excluded.
The regions are used to extract spectra (see text in \S 3).
``S" marks the region for the extended hard X-ray feature.
}
\end{figure}

\begin{figure}
\centerline{ {\hfil\hfil
\includegraphics[width=0.7\textwidth,angle=0, clip=]{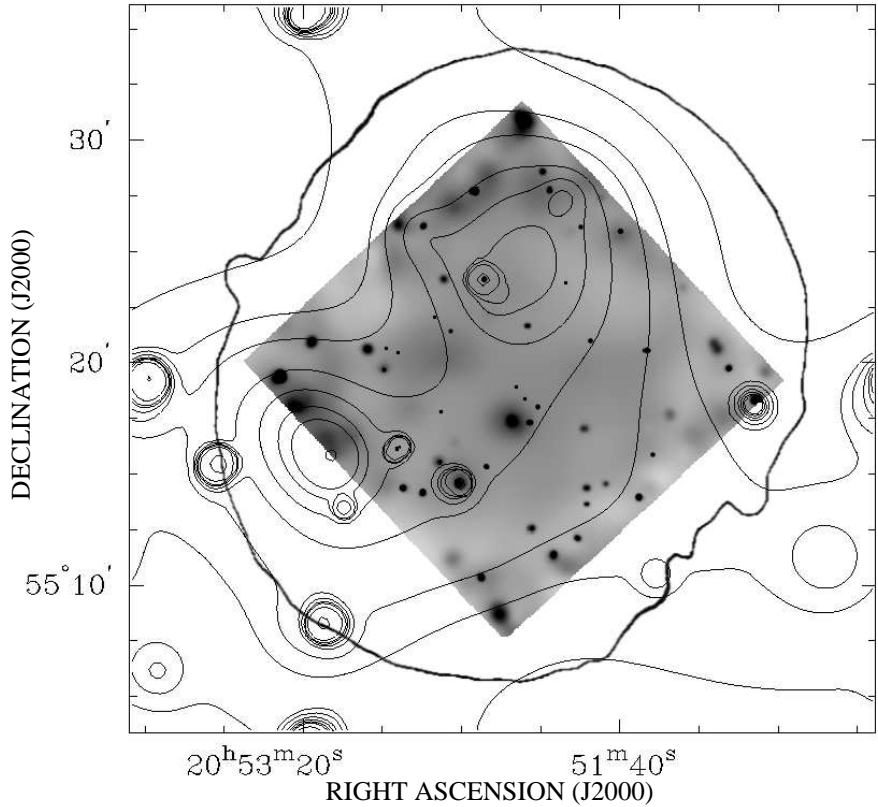}
\hfil\hfil}}
\caption{
\chandra\ ACIS-I intensity image of DA~530 in 0.5--8 keV band, compared with 
the intensity contours of the \rosat\ PSPC, which are 
spaced at 1.4, 1.65, 1.9, 2.45, 3, 3.6, 4.1, 5.9 and 30 
$\E{-5} \,{\rm photons}\cm^{-2}\ps\,\mbox{pixel}^{-1}$.
Both intensities have been exposure-corrected and smoothed using the 
CIAO tool {\em csmooth} to achieve a S/N ratio of 2.5--3.5. 
\chandra\ data are background-subtracted.
The same radio outer boundary (grey line) as in Fig.~1 is shown.
}
\end{figure}

\begin{figure}
\centerline{ {\hfil\hfil
\includegraphics[height=5.5in,angle=0, clip=]{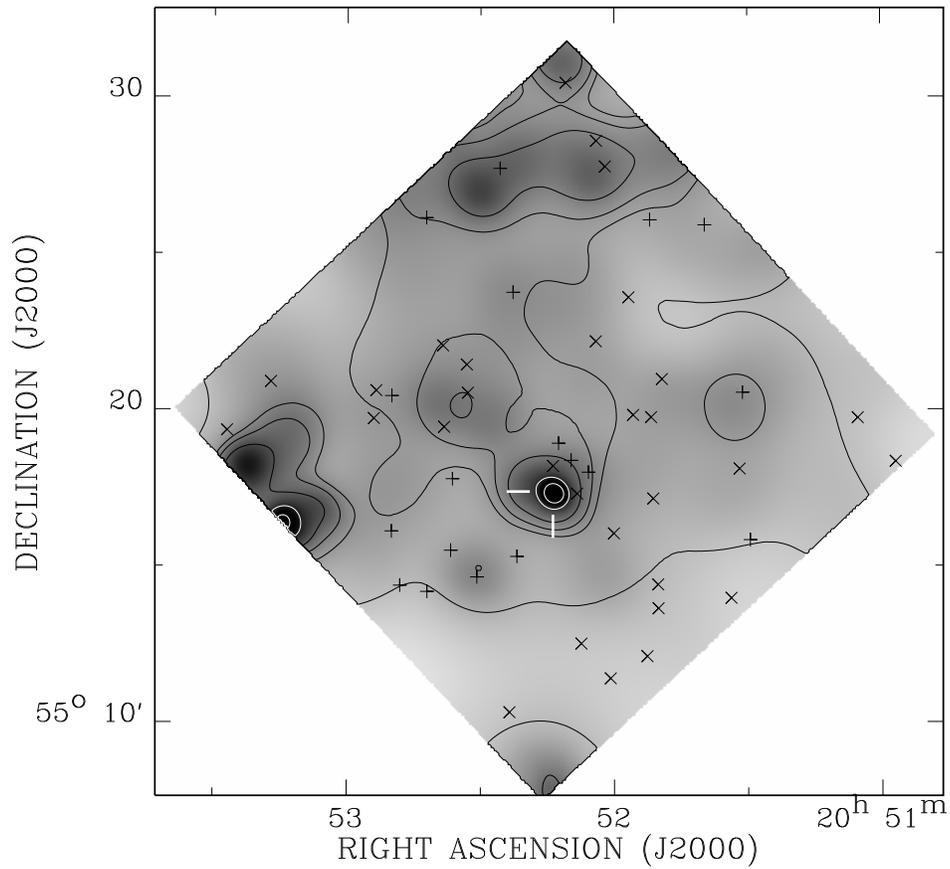}
\hfil\hfil}}
\caption{
Combined diffuse ACIS-I intensity images of DA~530 in 0.5--0.8 keV bands.
The contributions of the non-X-ray background and the detected point sources 
are subtracted. 
18 sources listed in Kaplan et~al.\ (2004) are marked with pluses 
as well as 30 new ones with crosses.
The superimposed 0.5--2 keV intensity contours are 
at 1.9, 2.19, 2.3, 2.5, 3.0 and 3.5 $\E{-3}$ cts $\cm^{-2}$s$^{-1}$ arcmin$^{-2}$.
The central bright extended feature is marked with two ticks.
}
\end{figure}

\begin{figure}
\centerline{ {\hfil\hfil
\includegraphics[height=4.0in,angle=0, clip=]{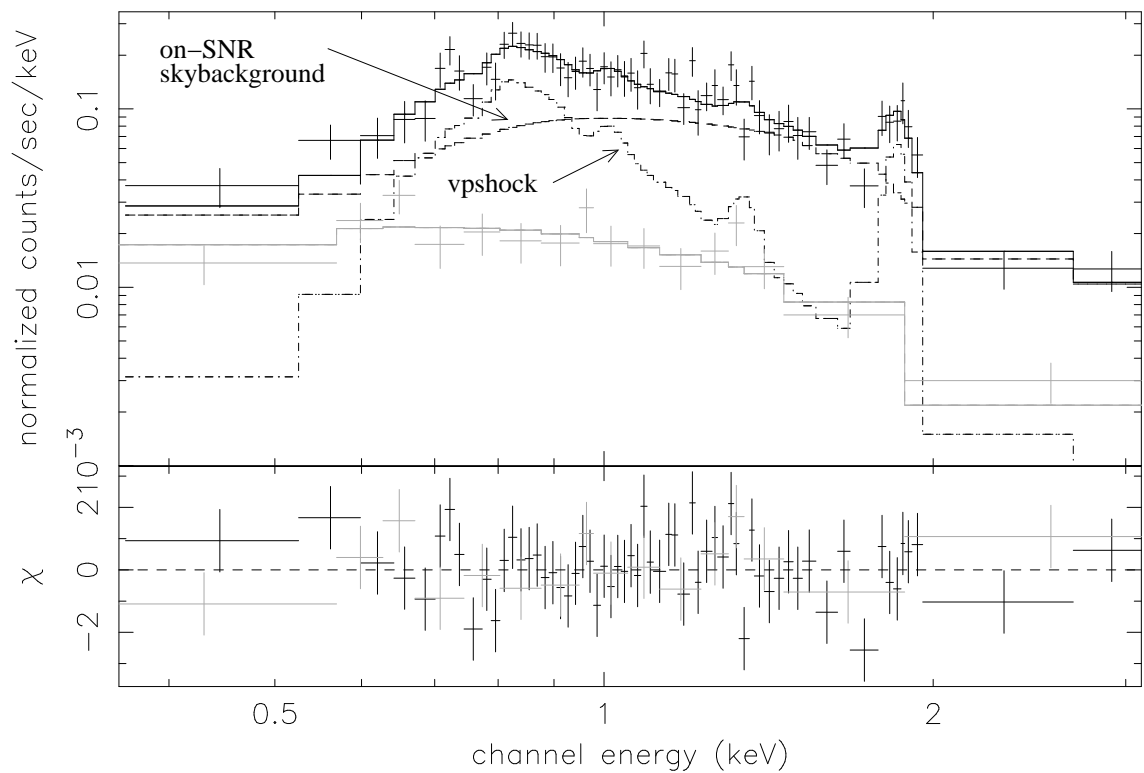}
\hfil\hfil}}
\caption{
The on-SNR spectra of DA~530 (black) and the local background (grey).
Also shown are the best-fit model of on-SNR spectrum (solid lines)
consisted of the free $\NH$ {\em vpshock} model (see Table~1; diffuse emission) 
and a power-law ($\Gamma\sim2$; sky background).
The two individual components are also plotted seperately with broken lines.
The residuals of the fit are shown in the lower panel.
}
\end{figure}

\begin{figure}
\centerline{ {\hfil\hfil
\includegraphics[height=5.5in,angle=0, clip=]{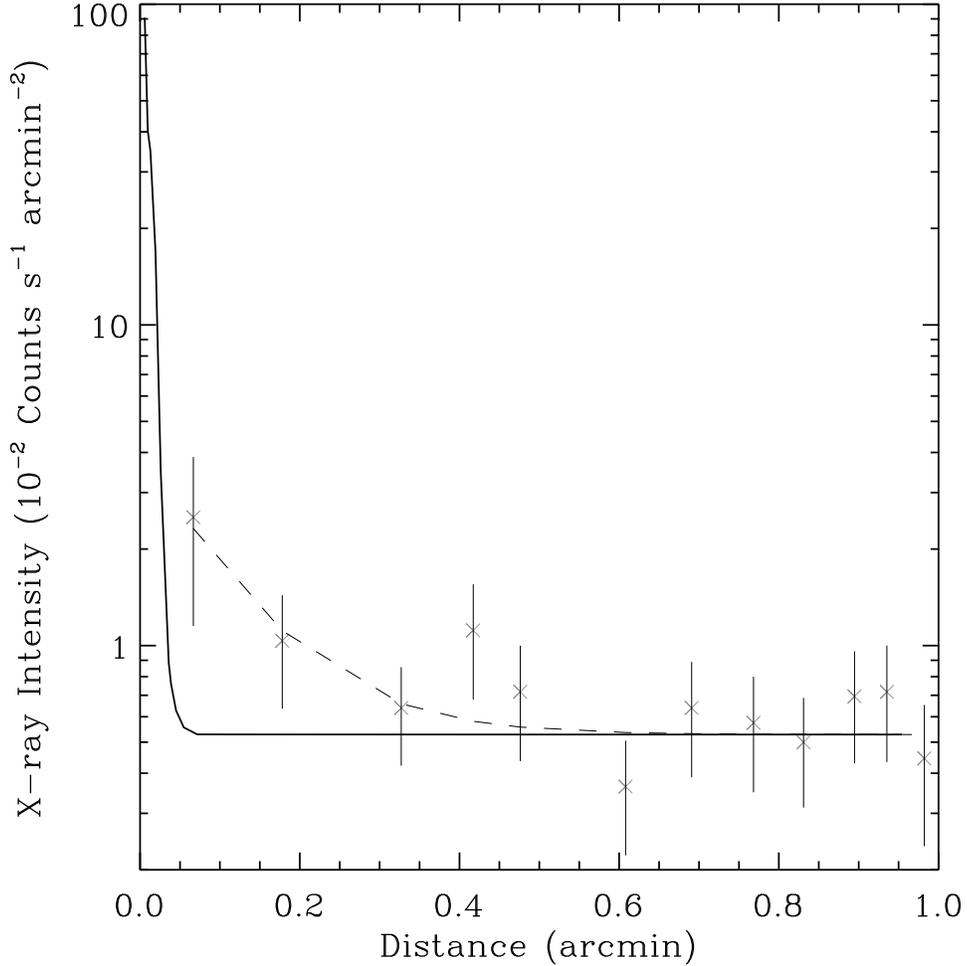}
\hfil\hfil}}
\caption{
The radial ACIS-I intensity profile of the central bright feature (crosses 
with error bars) compared with the ACIS-I 0.5-8 keV PSF (solid line) at the same 
position, to show the extension of the feature.  The data of 
the feature are exposure-corrected and are adaptively binned to achieve a 
count-to-noise ratio greater than 3 for the first point and 5 for outer 
points.
The PSF is simulated with 
{\sl MARX} (v4.2.1), normalized to the total net flux of the source and
presented together with the adopted local background estimated from the  
intensity profile in the $1'-1.5'$ annulus.
Dashed line represents a fit to the intensity profile with $I=0.035 e^{-r/0.1}+0.0053$. 
}
\end{figure}

\begin{figure}
\centerline{ {\hfil\hfil
\includegraphics[height=4.0in,angle=270, clip=]{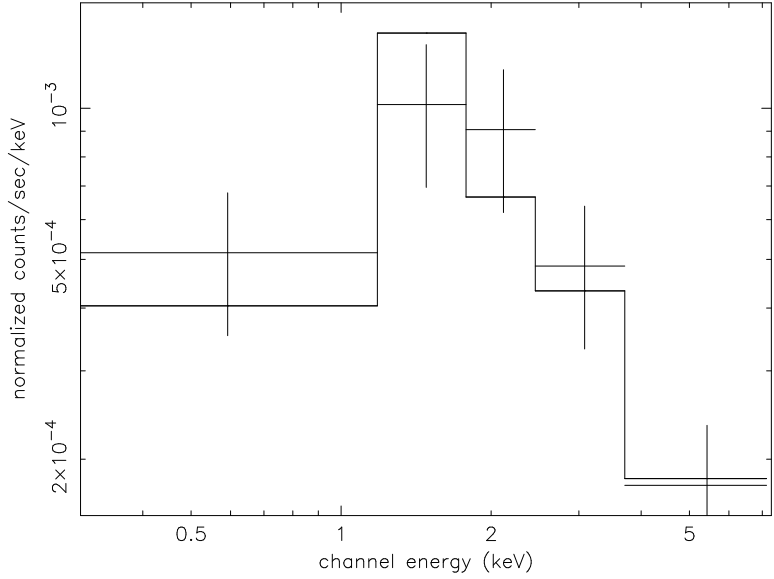}
\hfil\hfil}}
\caption{
The ACIS-I spectrum of the central extended feature, together with the 
best-fit power-law model with fixed $\NH=5.7\E{21}\cm^{-2}$ and $\Gamma=1.6$. 
The counts are grouped to have the count-to-noise ratio $>$ 3 per bin.
}
\end{figure}

\begin{figure}
\centerline{ {\hfil\hfil
\includegraphics[width=0.5\textwidth,angle=0, clip=]{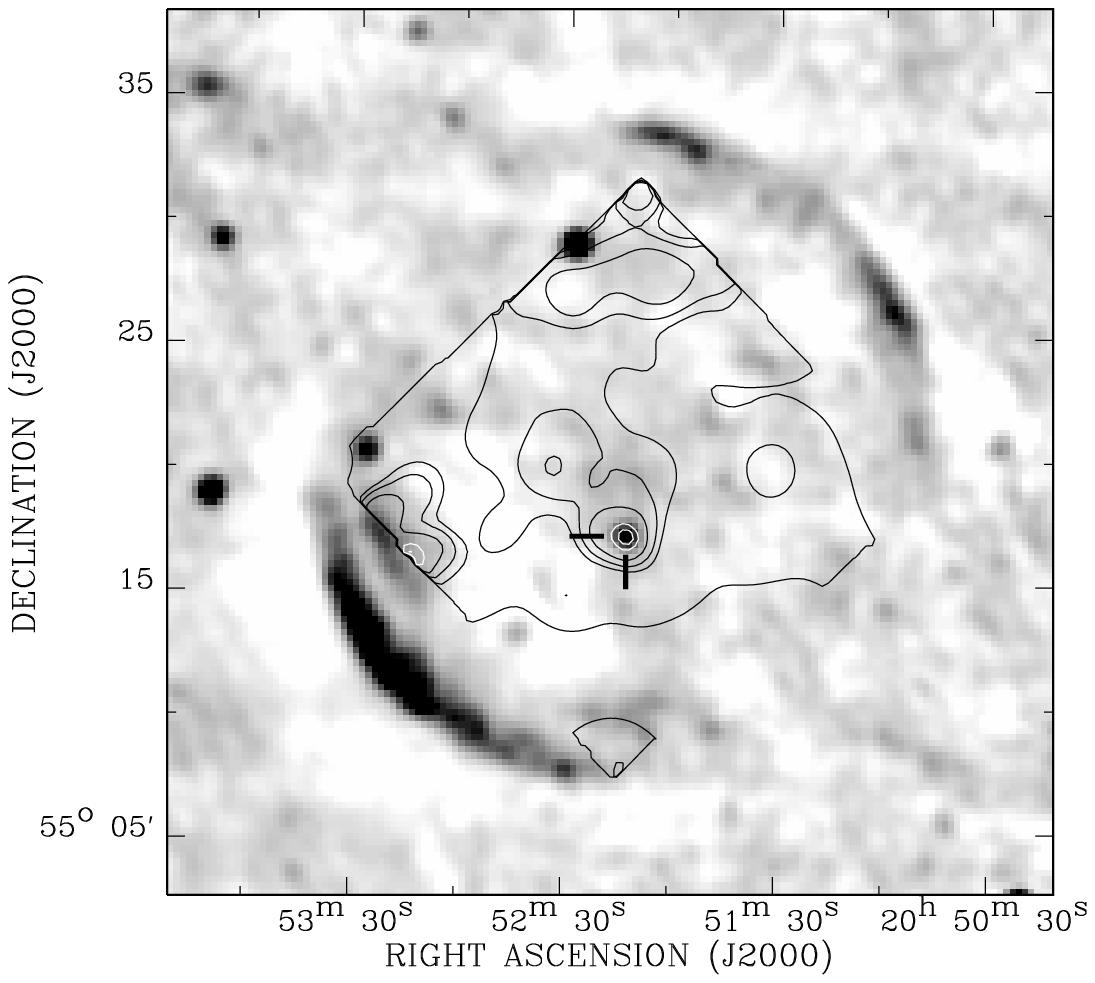}
\includegraphics[width=0.5\textwidth,angle=0, clip=]{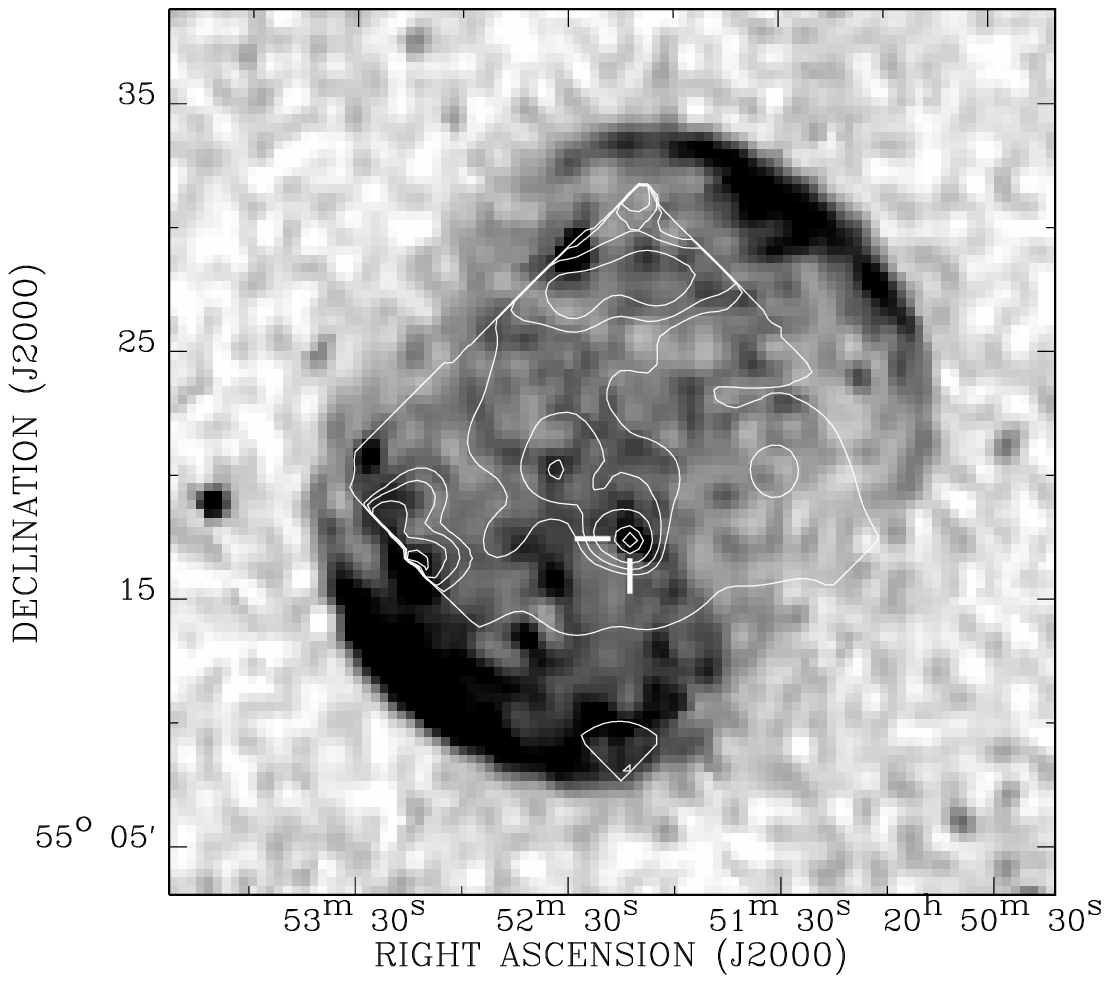}
\hfil\hfil}}
\caption{
{\em Left}: The 1.4 GHz NRAO VLA Sky Survey intensity map, {\em Right}: 
the 325~MHz Westerbork Northern Sky Survey intensity map. The scales are linear.
The same 0.5-2 keV intensity contours as in Fig.~3 except for the highest level, 
are overlaid to show 
the position correspondence of the PWN candidate (marked by two sticks) 
with the radio source.
}
\end{figure}

\end{document}